\documentclass[english,aps,prx,reprint,superscriptaddress]{revtex4-1}
\usepackage[T1]{fontenc}
\usepackage[latin9]{inputenc}
\usepackage{geometry}
\usepackage{color}
\usepackage{babel}
\usepackage{amsmath}
\usepackage{amssymb}
\usepackage{graphicx}
\usepackage[unicode=true,pdfusetitle,
 bookmarks=true,bookmarksnumbered=false,bookmarksopen=false,
 breaklinks=false,pdfborder={0 0 0},pdfborderstyle={},backref=false,colorlinks=true]
 {hyperref}
\hypersetup{
 linkcolor=blue, urlcolor=blue, citecolor=blue}

\makeatletter
\usepackage{braket}
\usepackage{dsfont}
\usepackage{bm}
\date{}

\makeatother

\begin{document}
\title{Negativity-Mutual Information conversion and coherence in two-coupled
harmonic oscillators}
\author{Jonas F. G. Santos}
\email{jonas.floriano@ufabc.edu.br}

\affiliation{Centro de Ciências Naturais e Humanas, Universidade Federal do ABC,
Avenida dos Estados 5001, 09210-580 Santo André, São Paulo, Brazil}
\author{Carlos H. S. Vieira}
\email{vieira.carlos@ufabc.edu.br}

\affiliation{Centro de Ciências Naturais e Humanas, Universidade Federal do ABC,
Avenida dos Estados 5001, 09210-580 Santo André, São Paulo, Brazil}
\author{Pedro R. Dieguez}
\email{dieguez.p@ufabc.edu.br}

\affiliation{Centro de Ciências Naturais e Humanas, Universidade Federal do ABC,
Avenida dos Estados 5001, 09210-580 Santo André, São Paulo, Brazil}
\begin{abstract}
Quantum information is a common topic of research in many areas of
quantum physics, such as quantum communication and quantum computation,
as well as quantum thermodynamics. It can be encoded in discrete or
continuous variable systems, with the appropriated formalism to treat
it generally depending on the quantum system to be chosen. For continuous
variable systems, it is convenient to employ a quasi-probability function
to represent quantum states and non-classical signatures. The Wigner
function is a special quasi-probability function because it allows
to describe a quantum system in the phase space very similar to the
classical one. In this work we consider some informational aspects
of two-mode continuous variables systems. In particular we illustrate
how the mutual information may be useful to transfer the non-Gaussianity
property between two states. Moreover, the coherence of a Gaussian
state is studied when the system is coupled to a thermal reservoir
but also with an extra single-mode Gaussian state. Our results may
be, in principle, investigated in trapped ions setups, where the two-mode
system can be encoded in the vibrational modes of the ion.
\end{abstract}
\date{\today}
\maketitle

\section{Introduction}

The complete ability of quantification and manipulation of quantum
information is a paramount challenge to design new micro and nano
devices based on quantum properties of systems. It is expected that
this fine control would have important impacts on quantum communication
and quantum cryptography \cite{Pleni02,Weekbrook2014,Pirandola2017,Pirandola2015},
quantum metrology \cite{Cirac2019,Adesso2017}, quantum thermodynamics
\cite{Camati2019,Kosloff2019,Shanhe2019}, and quantum computation
\cite{Zureck01,Supre2019,Silva2015}. These current and future implications
emphasize the importance in understanding quantum information in as
many points of view as possible, mainly because different experimental
platforms require an appropriated set of tools to manipulate information.
Encoding and processing information depend on the quantum system to
be considered. There are basically two broad classes of quantum systems
for these purposes, the two-level and continuous variable systems.
The first of them has a large numbers of studies, ranging from the
use of quantum information to reverse the direction of heat in quantum
thermodynamics \cite{Partovi2008,Jevtic2012,Serra2019}, to investigate
backflow of information in non-Markovian dynamics \cite{Maniscalco2018,Maniscalco2017,Camati2020}.
The latter, despite of some relevant works as, for instance, in squeezed
states \cite{Feig2019,Giovannetti01,Serafini2018} and in photon addition
and subtraction \cite{Su2019}, possesses relevant points to be addressed,
as the quality of squeezing sources and detectors, and quantum computation
architecture \cite{Adesso,Weedbrook2012}.

In what concerns the case of quantum information in two-level systems,
it is considered that the most adequate formalism is the wave function
or, in general cases, the density operator approach \cite{Petruccione01,Nielsen}.
On the other hand, for continuous variable quantum systems, such as
in quantum optics \cite{Quantumopticsbook}, trapped ions \cite{Wineland2003,Sage2019},
cold atoms \cite{Biedermann2015} and cavity QED \cite{Han2018,Werlang2018},
it is common to employ, besides the density operator, a quasi-probability
function to represent the state of some system \cite{Zachosbook}.
When dealing with the quantum to classical transition, the most suitable
of these functions is the Wigner function \cite{Wigner1932,Case2008},
which is represented in a phase space similarly to that of the classical
physics. The Wigner function has given important contribution in quantum
physics, for instance, in implementing teleportation of a quantum
gate \cite{Chou2018} and in general uncertainty relations in quantum
systems \cite{Santos2015,Santos2016,Bernardini01}. However, in many
cases it has been used more as a mean of visualization than to properly
describe the state of the system. There exist, nonetheless, the so-called
phase-space formalism of quantum mechanics, in which the Wigner function
is the state of some system and all the observables are complex functions
represented in the phase space.

The core of the phase-space formalism of quantum mechanics is the
Weyl tranform \cite{Case2008,Rosenbaum2006}, which converts a given
operator in a complex function. The Wigner function is then defined
as the Weyl transform of the density operator, for pure or mixed states.
The Moyal product \cite{Zachosbook} is also an important ingredient,
once it provides the appropriated inner product, in order to guarantee
the uncertainty relations of the quantum theory. Based on these main
elements, a formalism to treat quantum information, non-classicality
signatures, and coherence can be established. In this work, we study
some properties of two-mode continuous variables systems from the
point of view of quantum information. In particular, we investigated
the mutual information between two systems and show that it can be
employed to transfer the negativity of a number state to a Gaussian
one in a unitary dynamics. Moreover, we quantify the coherence of
a Gaussian state when the system is in contact with a thermal reservoir
but also with another harmonic oscillator. We verify the increase
of the coherence in some time interval during the dynamics and that
this effect is in accordance with a non-trivial behavior of the fidelity.

We note that these effects could be, in principle, experimentally
performed in trapped ions setups, where the two-mode system is encoded
in the vibrational models of the ions \cite{Twamley1997,Wineland1996,Zhong2018,Ortiz2017}.
On the other hand, the negativity of CV states are important in quantum
information, where they can be employed as resources in different
protocols \cite{Paris2018}.

This manuscript is organized as follows. In section \ref{sec:Basic-properties-of}
we provide the basic properties of the Wigner function and the phase-space
formalism of quantum mechanics. In section \ref{sec:Evolution-of-a}
we consider a two-mode state evolving unitarily in order to show that
the mutual information allows the transference of the negativity of
a number state to a Gaussian one. Section \ref{sec:Dissipative-dynamics}
is devoted to study the dissipative dynamics of a one-mode Gaussian
state interacting with a thermal reservoir as well as with another
harmonic oscillator, from the point of view of the coherence during
the dynamics. Here it is shown that the fidelity works as a witness
of non-Markovian like-effect. We briefly discuss a trapped ions approach
that could be useful to implement our findings in Section \ref{sec:Trapped-ions-approach}.
Finally, we draw our conclusions and final remarks in section \ref{sec:Conclusions}.

\section{Theoretical framework\label{sec:Basic-properties-of}}

In this section we review some relevant properties of the Wigner function
and the phase-space formalism of quantum mechanics. The standard formalism
of quantum mechanics is based on operators that act on the Hilbert
space, $\mathcal{H}$. A quantum system can be described by means
of a wave function in the position representation or, through the
Fourier transform, in the momentum representation. Moreover, there
is the well-known commutation relation between position and momentum
operators, $[\hat{q}_{i},\hat{p}_{j}]=i\hbar\delta_{ij}$, where $i$
and $j$ run over all the Hilbert space.

The operators-based formalism is not the only way to describe quantum
mechanics. The phase-space formalism of quantum mechanics (PSQM) is
another interesting way to study quantum mechanical systems, being
relevant in a large number of scenarios. In the PSQM, the Wigner function
represents the state of a given system and has the important property
of carrying out simultaneously information about position and momentum
of the system. In order to introduce the Wigner function and the phase-space
formalism, we consider the Weyl transform of an operator $\hat{\mathcal{O}}(\hat{q},\hat{p})$,
defined as \cite{Zachosbook,Case2008},

\begin{equation}
A^{W}(q,p)=\int dy\,e^{-ipy/\hbar}\langle q+y/2|\hat{\mathcal{O}}(\hat{q},\hat{p})|q-y/2\rangle,\label{Weyl}
\end{equation}
where $W$ stands for Weyl transform. The Weyl transform converts
an operator in a c-function and it is a natural connection between
operators-based and phase-space formalisms.

Next, considering a system described by a pure state, $|\psi\rangle$,
we can write the density operator, $\hat{\rho}=|\psi\rangle\langle\psi|$.
The Wigner function can be defined just as the Weyl tranform of the
density operator,

\begin{align}
W(q,p) & =h^{-1}\int dy\,e^{-ipy/\hbar}\langle q+y/2|\hat{\mathcal{\rho}}|q-y/2\rangle,\nonumber \\
 & =h^{-1}\int dy\,e^{-ipy/\hbar}|\psi(q+y/2)\rangle\langle\psi(q-y/2)|.\label{wigner01}
\end{align}

Some important properties of the Wigner function are that it is normalized
when integrated over all phase-space, additionally providing the marginal
probability distribution for momentum $|\phi(p)|^{2}$, and position
$|\psi(q)|^{2}$,

\begin{equation}
\int dq\,W(q,p)=|\phi(p)|^{2},\quad\int dp\,W(q,p)=|\psi(q)|^{2}.\label{marginal}
\end{equation}

From the Wigner function, it is also possible to obtain the expectation
the value of an observable $\mathcal{O}$,

\begin{equation}
\langle\mathcal{O}\rangle=h^{-1}\int\int dqdp\,W(q,p)\mathcal{O}^{W}(q,p).\label{expectationvalue}
\end{equation}

The generalization of the Wigner function for mixed states is straightforward.
For a mixed state $\rho=\sum_{j}p_{j}|\psi_{j}\rangle\langle\psi_{j}|$,
where $p_{j}$ is the probability of each state and it is always positive,
with $\sum_{j}p_{j}=1$, the Wigner function is simply given by,

\begin{equation}
W(q,p)=\sum_{j}p_{j}W_{j}(q,p),\label{wigner02}
\end{equation}
with $W_{j}(q,p)$ the Wigner function associated to each part of
the ensemble.

One of the basic ingredients in considering the PSQM formalism is
the Moyal product, which is introduced once we are now dealing with
c-functions and no longer with operators. The Moyal product, defined
for position and momentum variables \cite{Zachosbook}, reads $\mathcal{A}^{W}(q,p)\star\mathcal{B}^{W}(q,p),$
where,

\begin{equation}
\star=\text{exp}\left[\frac{i\hbar}{2}\left(\frac{\overleftarrow{\partial}}{\partial q}\frac{\overrightarrow{\partial}}{\partial p}-\frac{\overleftarrow{\partial}}{\partial p}\frac{\overrightarrow{\partial}}{\partial q}\right)\right].\label{moyal}
\end{equation}

In introducing the Moyal product into the PSQM formalism we obtain
a set of important tools to treat quantum systems, most of them being
very similar to the classical case. In particular, for a unitary dynamics,
the time evolution of an observable $\mathcal{O}^{W}$is dictated
by the equation,

\begin{align}
\dot{\mathcal{O}}^{W}(q,p;t) & =-\frac{i}{\hbar}\left[\mathcal{O}^{W}(q,p;0),H^{W}(q,p;t)\right]_{\star}\nonumber \\
 & =\frac{i}{\hbar}\mathcal{O}^{W}(q,p;0)\star H^{W}(q,p;t)\label{motioneq}\\
 & -\frac{i}{\hbar}H^{W}(q,p;t)\star\mathcal{O}^{W}(q,p;0),
\end{align}
well known as Moyal equation. For a generalization of the Moyal equation
for open quantum systems we refer to Ref. \cite{Deering2015}.

In addition, applying the Weyl tranform on the eigenvalue equation
and using the Moyal product, we obtain the so-called stargenvalue
equation, given by \cite{Zachosbook,Rosenbaum2006},

\begin{equation}
H^{W}(q,p)\star W(q,p)=EW(q,p),\label{eigenvalueeq}
\end{equation}
where $H^{W}(q,p)$ is the Weyl tranform of the Hamiltonian of a quantum
system and $E$ sets for the eigenvalues of energy. In the case of
a quantum system with $N$ dimensions, the selection of a particular
sub phase space $(q_{k},p_{k})$ is performed by integrating over
the rest of the variables,

\begin{equation}
W(q_{k},p_{k})=\int\int dq_{N-k}dp_{N-k}\,W(q_{\ell},p_{\ell}).\label{trace}
\end{equation}

Additionally, we present how to obtain the time evolution of the Wigner
function using the Moyal product in the case of a unitary operation.
Once the initial Wigner function is obtained from Eq. (\ref{eigenvalueeq}),
the time evolution is given by the unitary operator \cite{Zachosbook},

\begin{align}
U_{\star}(q_{\ell},p_{\ell};t) & =e^{itH^{W}/\hbar}=1+(it/\hbar)H^{W}\label{evolution}\\
 & +\frac{(it/\hbar)^{2}}{2!}H^{W}\star H^{W}+...,
\end{align}
 with $\ell=1,...,N,$and $N$ the dimension of the system, resulting
in,

\begin{equation}
W(q_{\ell},p_{\ell};t)=U_{\star}^{-1}(q_{\ell},p_{\ell};t)\star W(q_{\ell},p_{\ell};0)\star U_{\star}(q_{\ell},p_{\ell};t).\label{timewigner}
\end{equation}

Finally, for the case in which the system is a two-mode Gaussian state,
we can introduce a vector collecting all the coordinates of the phase
space, $\vec{R}(q_{1},p_{1},q_{2},p_{2})$. The two-mode Gaussian
state is then completely characterized in terms of its first moments
and the covariance matrix, defined as $\vec{d}=(\langle q_{1}\rangle,\langle p_{1}\rangle,\langle q_{2}\rangle,\langle p_{2}\rangle)$
and $\sigma=\sigma_{11}\oplus\sigma_{22}$ , respectively, with,

\begin{equation}
\sigma_{ii}=\left(\begin{array}{cc}
\sigma_{q_{i}q_{i}} & \sigma_{p_{i}q_{i}}\\
\sigma_{q_{i}p_{i}} & \sigma_{p_{i}p_{i}}
\end{array}\right),\label{CM}
\end{equation}
and $\sigma_{AB}=\langle AB+BA\rangle-2\langle A\rangle\langle B\rangle$.
This results in a convenient expression for obtaining the Wigner function
\cite{Adesso,Weedbrook2012},

\begin{equation}
W_{G}(\vec{R})=\frac{exp\left[-(1/2)(\vec{R}-\vec{d})\sigma^{-1}(\vec{R}-\vec{d})\right]}{(2\pi)^{4}\sqrt{Det[\sigma]}},\label{wigner}
\end{equation}
 particularizing the expression for two modes.

Now, we present some informational quantifiers which will be relevant
for our discussion. The first point to be addressed is the introduction
of linear entropy. The treatment will be restricted for a bipartite
system. Since a direct translation of the von Neumann entropy to the
phase-space is not generally possible because the Wigner function
may assume negative values, the main alternative adopted is to employ
the well-known linear entropy, defined as \cite{Mandredi2000},

\begin{equation}
S=1-(2\pi\hbar)^{2}\int\int dq_{i}dp_{i}\,W^{2}(q_{i},p_{i}),\label{linearentropy}
\end{equation}
with $i=1,2$.

The Eq. (\ref{linearentropy}) holds for pure and mixed quantum states
and satisfies the relation $0\leq S\leq1$, where $S=0$ for a pure
state.

\textbf{Mutual information. }The linear entropy is useful to define
the mutual information between two parts of a bipartite quantum system.
Let consider a quantum system described by a Wigner function $W(q_{i},p_{i})$
where the parts may have some physical interaction. The mutual information
shared between the two parts is defined as \cite{Santos2015,Bernardini01,Mandredi2000},

\begin{equation}
I^{(1:2)}=S(W_{1})+S(W_{2})-S(W),\label{mutual}
\end{equation}
where $W_{1}$ and $W_{2}$ are the local Wigner functions of each
part,

\begin{align*}
W_{1} & =W(q_{1},p_{1})=\int\int dq_{2}dp_{2}\,W(q_{i},p_{i}),\\
W_{2} & =W(q_{2},p_{2})=\int\int dq_{1}dp_{1}\,W(q_{i},p_{i}).
\end{align*}

The mutual information measures the total correlation, classical or
quantum, between two parts of a system. When the bipartite system
is separable, $W(q_{i},p_{i})=W(q_{1},p_{1})W(q_{2},p_{2})$, the
mutual information is zero.

\textbf{Wigner function negativity. }Another way of quantifying non-classical
effects in a system is the negativity of the local Wigner function,
defined in a bipartite system as \cite{Kenfack2014},

\begin{align*}
\delta(W_{1}) & =\int\int dq_{1}dp_{1}\,|W_{1}|-\int\int dq_{1}dp_{1}\,W_{1},\\
\delta(W_{2}) & =\int\int dq_{2}dp_{2}\,|W_{2}|-\int\int dq_{2}dp_{2}\,W_{2}.
\end{align*}

The negativity of the Wigner function has been used in different cases
for investigating non-classical effects \cite{Rundle2017,Treps2017,Svozil=0000EDk2018}
and reported experimentally in Ref. \cite{Haroche2000}, with applications
in quantum computation \cite{Okay2014}. For separable or correlated
bipartite states the negativity is directly related to non-classical
effects of each part. Thus, using the mutual information and the negativity
it is possible to have a broad knowledge on the correlations shared
between the two parts of a system.

\subsection*{Particular case: Gaussian states}

Here we detail some quantum information quantities in the particular
case in which the Wigner function is Gaussian, i.e. completely characterized
by the first and second moments \cite{Adesso,Serafini,Hiroshima2007,Weedbrook2012}.

\textbf{Fidelity.} The fidelity is a suitable way of comparing how
similar (or distinct) two states are. Consider two Gaussian states,
characterized by their first and seconds moments (covariance matrix),
$\vec{d}$ and $\sigma$, respectively, such that we can write the
states as $W_{1}(\vec{d}_{1},\sigma_{1})$ and $W_{2}(\vec{d}_{2},\sigma_{2})$.
The fidelity is given by \cite{Holevo,Scutaru},

\begin{equation}
F(\sigma_{1},\vec{d}_{1};\sigma_{2},\vec{d}_{2})=\frac{2}{\sqrt{\Delta+\delta}-\sqrt{\delta}}e^{-\frac{1}{2}\vec{d}^{T}\sigma_{+}^{-1}\vec{d}},\label{fidelity}
\end{equation}
where $\Delta\equiv Det[\sigma_{1}+\sigma_{2}]$, $\delta=(Det[\sigma_{1}]-1)(Det[\sigma_{2}]-1)$,
$\vec{d}\equiv\vec{d}_{1}-\vec{d}_{2}$, and $\sigma_{+}=\sigma+\sigma_{2}$.
The fidelity is bounded by $0\leq F\leq1$, with $F=1$ and $F=0$
for identical and completely different states, respectively. Equation
(\ref{fidelity}) has particular importance in quantum information
processing with Gaussian states because it allows, for example, to
quantify the quality of a given noisy communication channel, where
we have Gaussian states as input and output \cite{Hiroshima2007}.
Another important point is that the covariance matrix is experimentally
accessible, in particular in quantum optical devices \cite{Hage2007}.

\textbf{Coherence.} An important quantum signature is the presence
of quantum coherence in states. This quantity has been addressed in
a series of recent articles in quantum information \cite{Winter2016,Plenio2014},
and also in quantum thermodynamics \cite{Camati2019,Feldmann2006,Rezek2006}.
The quantification of coherence as a resource was firstly proposed
in Ref. \cite{Plenio2014} and extended for Gaussian states in Ref.
\cite{Xu2016}. In order to quantify the coherence in a given one-mode
Gaussian state $\rho(\sigma,\vec{d})$, where $\sigma$ and $\vec{d}$
are the covariance matrix and the first moments, respectively, Ref.
\cite{Xu2016} introduces a quantifier defined as $C[\rho(\sigma,\vec{d})]=\text{min}\left\{ S[\rho(\sigma,\vec{d})||\rho^{th}]\right\} $,
where $S[\bullet||\bullet]$ is the relative entropy and the minimum
is evaluated over all thermal states $\rho^{th}$. The coherence quantifier
for one-mode Gaussian states assumes the following expression when
minimized \cite{Xu2016},

\begin{align}
C[\rho(\sigma,\vec{d})] & =\frac{\nu-1}{2}log_{2}\left(\frac{\nu-1}{2}\right)-\frac{\nu+1}{2}log_{2}\left(\frac{\nu+1}{2}\right)\nonumber \\
 & +(\bar{n}+1)log_{2}\left(\bar{n}+1\right)-(\bar{n})log_{2}\left(\bar{n}\right)\label{coherence}\\
 & =S(\rho^{th})-S(\rho),
\end{align}
where $\nu=\sqrt{\sigma_{11}\sigma_{22}-\sigma_{12}^{2}}$, and $\bar{n}=(1/4)(\sigma_{11}+\sigma_{22}+d_{1}^{2}+d_{2}^{2}-2)$.
As we will observe in the following examples, this quantity is appropriated
to capture the coherence of Gaussian states due to the displacement
operator.

\textbf{Dissipative dynamics of Gaussian states.} When a system is
in thermal contact with a Markovian environment, in general it is
affected by decoherence effects \cite{Petruccione01}. In the particular
case when the system is a one-mode Gaussian state $\rho(\sigma,\vec{d})$
the complete characterization of dissipation effects is encoded into
the dynamics of the first moments and the covariance matrix \cite{Serafini,Giovanneti01}.
The time evolution of the first moments and the covariance matrix
during the thermalization process is obtained from the two uncoupled
differential equations,

\begin{eqnarray*}
\dot{\sigma} & = & \Gamma\sigma+\Gamma(2\bar{m}+1)\mathbb{I}_{2\times2},\\
\dot{\vec{d}} & = & -(\Gamma/2)\vec{d},
\end{eqnarray*}
where $\Gamma$ and $\bar{m}$ are the decay rate and the mean number
of photons of the thermal environment, respectively. The solutions
for the above equations are straightforward and given by,

\begin{align}
\sigma(t) & =e^{-\Gamma t}\sigma(0)+(1-e^{-\Gamma t})(2\bar{m}+1)\mathbb{I}_{2\times2},\\
\vec{d}(t) & =e^{-\Gamma t/2}\vec{d}(0),
\end{align}
with $\vec{d}(0)$ and $\sigma(0)$ the initial first moments and
covariance matrix of the system. Naturally, when $t\rightarrow\infty$,
the first moments and the covariance matrix tend to the asymptotic
thermal state.

\section{Negativity transference due to mutual information\label{sec:Evolution-of-a}}

In this section we consider a unitary dynamics of two correlated systems
and show that the mutual information works as a kind of resource to
allow the transference of negativity. We assume the following Hamiltonian,

\begin{equation}
H(q_{i},p_{i})=\alpha^{2}p_{i}+\beta^{2}q_{i}^{2}+\gamma(p_{1}q_{2}-p_{2}q_{1}),\label{Hcoupled}
\end{equation}
with $\alpha^{2}=1/(2m)$ and $\beta^{2}=m\omega^{2}/2$, where $m$
and $\omega$ are the mass and frequency of the system, respectively,
and $\gamma$ represents the coupling constant between the two oscillators.

The Hamiltonian (\ref{Hcoupled}) illustrates many important scenarios
in physics, ranging from fluctuation relations for a particle in a
magnetic field \cite{Sahoo2007} to general uncertainty relations
\cite{Rosenbaum2006,Bernardini01,Santos2015,Santos2016} and quantum
heat engines \cite{Munos2014,Santos2017}. For example, Eq. (\ref{Hcoupled})
could describe a particle moving in a plane in the presence of an
orthogonal uniform magnetic field. Another important scenario where
this Hamiltonian could be interesting is in trapped ions setup, where
the vibrational modes of the ion is described by continuous variables
systems \textbackslash cite{[}Generation and stabilization of entangled
coherent states for the vibrational modes of a trapped ion{]}. In
\ref{App} we provide a detailed derivation of the Wigner function
and associated eigenvalues for the Hamiltonian (\ref{Hcoupled}) using
Eq. (\ref{eigenvalueeq}). Denoting by $W_{n_{1},n_{2}}(q_{i},p_{i})$
the Wigner function with eigenvalues $E_{n_{1},n_{2}}$ reads

\begin{align}
W_{n_{1},n_{2}}(q_{i},p_{i}) & =\frac{(-1)^{n_{1}+n_{2}}}{\pi^{2}\hbar^{2}}\text{exp}\left[-\frac{1}{\hbar}\left(\frac{\alpha}{\beta}q_{i}^{2}+\frac{\beta}{\alpha}p_{i}^{2}\right)\right]\nonumber \\
 & \times L_{n_{1}}[\Omega_{+}/\hbar]L_{n_{2}}[\Omega_{-}/\hbar],\label{WignerA}
\end{align}
with $\Omega_{\pm}=(\alpha/\beta)q_{i}^{2}+(\beta/\alpha)p_{i}^{2}\mp2\sum_{i,j=1}^{2}(\epsilon_{ij}p_{i}q_{j})$,
$n_{1(2)}$ are interger and nonnegative numbers, $L_{n_{1}(n_{2})}$
are the associated Laguerre polynomials, and the eigenvalues are,

\begin{equation}
E_{n_{1},n_{2}}=2\hbar\alpha\beta(n_{1}+n_{2}+1)+\hbar\gamma(n_{1}-n_{2}).\label{Ea}
\end{equation}

For our purpose the dynamics of the whole system is unitary and it
is obtained by evaluating the Eq. (\ref{motioneq}), which results
in set of uncoupled four equations of motions for $H(q_{i},p_{i})$,

\begin{align}
q_{1}(t) & =x_{0}\cos(\omega t)\cos(\gamma t)+y_{0}\cos(\omega t)\sin(\gamma t)\nonumber \\
 & +\frac{\beta}{\alpha}[p_{y_{0}}\sin(\omega t)\sin(\gamma t)+p_{x_{0}}\sin(\omega t)\cos(\gamma t)],\label{sol1}\\
q_{2}(t) & =y_{0}\cos(\omega t)\cos(\gamma t)-x_{0}\cos(\omega t)\sin(\gamma t)\nonumber \\
 & -\frac{\beta}{\alpha}[p_{x_{0}}\sin(\omega t)\sin(\gamma t)-p_{y_{0}}\sin(\omega t)\cos(\gamma t)],\label{sol2}\\
p_{1}(t) & =p_{x_{0}}\cos(\omega t)\cos(\gamma t)+p_{y_{0}}\cos(\omega t)\sin(\gamma t)\nonumber \\
 & -\frac{\alpha}{\beta}[y\sin(\omega t)\sin(\gamma t)+x\sin(\omega t)\cos(\gamma t)],\label{sol3}\\
p_{2}(t) & =p_{y_{0}}\cos(\omega t)\cos(\gamma t)-p_{x_{0}}\cos(\omega t)\sin(\gamma t)\nonumber \\
 & +\frac{\alpha}{\beta}[x_{0}\sin(\omega t)\sin(\gamma t)-y_{0}\sin(\omega t)\cos(\gamma t)],\label{sol4}
\end{align}
where $x_{0}$, $y_{0}$, $p_{x_{0}}$, and $p_{y_{0}}$ are arbitrary
initial parameters.

The Wigner function in Eq. (\ref{WignerA}) are clearly stationary.
Based on Ref. \cite{Bernardini01}, we can write an initial state
$W_{k,\ell}(q_{i},p_{i})$ which are stationary only for the particular
case of $\gamma=0$ and is time dependent for any other value of $\gamma$.
Following Ref. \cite{Bernardini01}, this state reads,

\begin{equation}
W_{k,\ell}(q_{i},p_{i})=\frac{(-1)^{k+\ell}}{\pi^{2}\hbar^{2}}\exp\left[-\xi_{1(2)}^{2}/\hbar\right]L_{k(\ell)}^{(0)}\left[2\xi_{1(2)}^{2}/\hbar\right],\label{Wigner11}
\end{equation}
 with

\begin{align*}
\xi_{1}^{2} & =\left(\frac{\alpha}{\beta}q_{1}^{2}+\frac{\beta}{\alpha}p_{1}^{2}\right)\cos(\gamma t)^{2}+\left(\frac{\alpha}{\beta}q_{2}^{2}+\frac{\beta}{\alpha}p_{2}^{2}\right)\sin(\gamma t)^{2}\\
 & -\left(\frac{\alpha}{\beta}q_{1}q_{2}+\frac{\beta}{\alpha}p_{1}p_{2}\right)\sin(2\gamma t),\\
\xi_{2}^{2} & =\left(\frac{\alpha}{\beta}q_{1}^{2}+\frac{\beta}{\alpha}p_{1}^{2}\right)\sin(\gamma t)^{2}+\left(\frac{\alpha}{\beta}q_{2}^{2}+\frac{\beta}{\alpha}p_{2}^{2}\right)\cos(\gamma t)^{2}\\
 & +\left(\frac{\alpha}{\beta}q_{1}q_{2}+\frac{\beta}{\alpha}p_{1}p_{2}\right)\sin(2\gamma t).
\end{align*}

\begin{figure}
\includegraphics[scale=0.75]{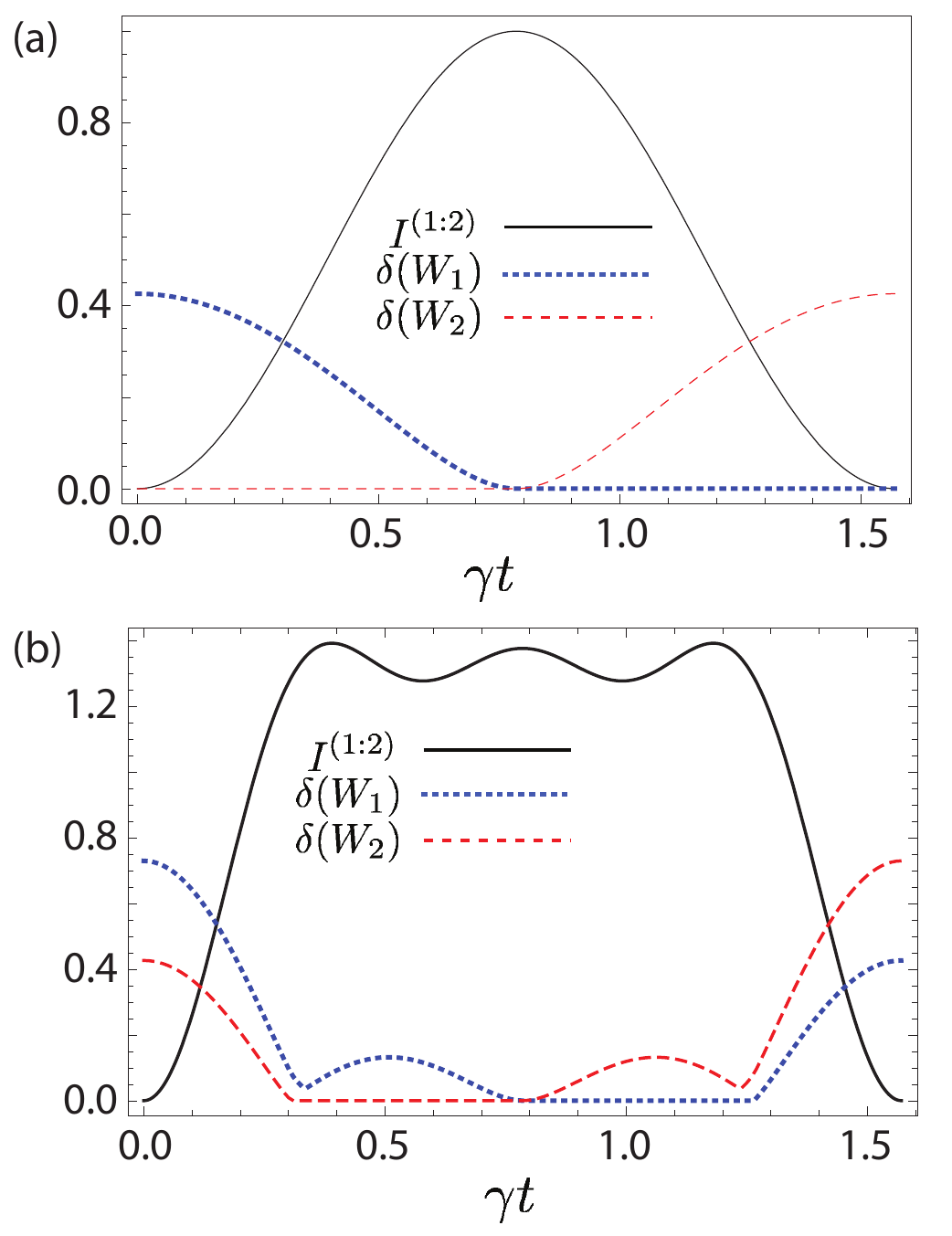}

\caption{(Color online): Mutual information (black solid line) and negativity
of the local Wigner functions $W_{1}(q_{1},p_{1})$ (blue dotted line)
and $W_{2}(q_{2},p_{2})$ (red dashed line ) as function of the dimensionless
time $\gamma t$ for two different cases, (a) $(k,\ell)=(1,0)$ and
(b) $(k,\ell)=(2,1)$, in order to illustrate how the mutual information
depends on the non-classicality of local states of the system. It
is possible to note the inversion of the negativities in both cases,
as well as that the minimum value of them is reached when the mutual
information is maximum. We considered unities such that $\hbar=m=\omega=1$.}

\label{Grafico01}
\end{figure}

Figure (\ref{Grafico01}) shows the mutual information (black solid
line) and the negativity of the local Wigner functions for two sets
of quantum numbers, $(k,\ell)=(1,0)$ and $(k,\ell)=(2,1)$. In the
first case, Fig. (\ref{Grafico01})-(a), we observe that the negativity
of $W_{1}(q_{1},p_{1})$ (blue dotted line) and $W_{2}(q_{2},p_{2})$
(red dashed line) are initially non-zero and zero values, respectively,
as expected, once these states are the first and the ground (Gaussian)
states of the harmonic oscillator. As the time evolves, the mutual
information shared between the subsystems $(q_{1},p_{1})$ and $(q_{2},p_{2})$
increases up to the negativity of $W_{1}(q_{1},p_{1})$ reaches the
minimum value, resulting in a decrease of the negativity of $W_{1}(q_{1},p_{1})$.
Then, the mutual information starts to decrease while the negativity
of $W_{2}(q_{2},p_{2})$ increases up to the maximum value. In the
second case, Fig. (\ref{Grafico01})-(b), we have two non-Gaussian
states initially with non-zero initial values of negativity of the
local Wigner functions. Similarly to the first case, the negativities
decrease as the increase of the mutual information and, after a period
of time, the mutual information decreases while the negativies increases
with inverse values. In both cases, we note the negativity-mutual
information conversion, highlighting the role of the negativity in
supplying correlations. Our results can be, in principle, experimentally
implemented in continuous variable platforms, such as in specific
ions trap in which there are two vibrational degrees of freedom accessible
\cite{Twamley1997}.

\section{Dissipative dynamics \label{sec:Dissipative-dynamics}}

\begin{figure}
\includegraphics{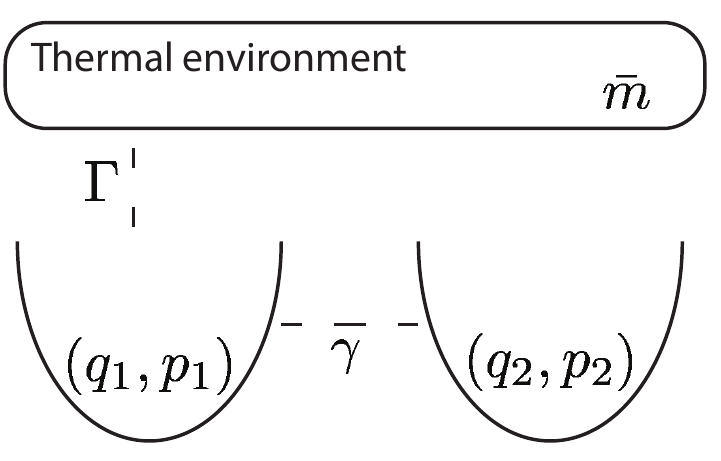}

\caption{Illustration of the system which is considered, $W_{sys}=W_{1}(q_{1},p_{1})$,
coupled to another quantum oscillator $(q_{2},p_{2})$ by the coupling
$\gamma$, and also coupled to a thermal environment with coupling
$\Gamma$. We are considering that the coupling between the thermal
environment and the quantum oscillator $(q_{2},p_{2})$ is sufficiently
weak such that any effect due to it is negligible for the evolution
of the system $W_{sys}$.}

\label{illustration}
\end{figure}

For this purpose, we shall assume that the states of the subsystems
$(q_{1},p_{1})$ and $(q_{2},p_{2})$ are Gaussian, i.e. $W_{1}(q_{1},p_{1})$
and $W_{2}(q_{2},p_{2})$ are completely characterized by their first
moments and covariance matrix, and defining our system as being $W_{sys}\equiv W_{1}(q_{1},p_{1})$.
Besides, we consider that $\Gamma$ and $\gamma$ are constants mediating
the coupling between the thermal reservoir with the subsystem $(q_{1},p_{1})$,
and the subsystems $(q_{1},p_{1})$ and $(q_{2},p_{2})$, respectively.
Figure (\ref{illustration}) shows an illustration of the considered
dissipative process. We restrict our attention to investigate the
dynamics of $W_{sys}$ and observe how the coupling $\gamma$ impacts
its dissipative dynamics when in thermal contact with a Markovian
environment. From a physical point of view, we consider that the time
evolution of $W_{sys}$ is strictly Markovian when $\gamma\rightarrow0$.
Moreover, in tracing out the degrees of freedom $(q_{2},p_{2})$ and
restricting our attention only to $W_{sys}$, we are assuming that
the coupling between the thermal environment and the degrees of freedom
$(q_{2},p_{2})$ is sufficiently weak such that it does not cause
any effect on the evolution of $W_{sys}$. 

\begin{figure}
\includegraphics[scale=1.9]{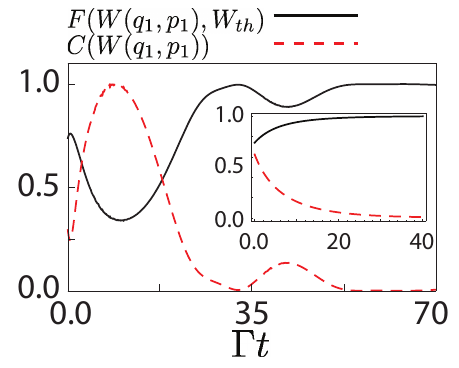}

\caption{(Color online): Fidelity and normalized coherence for the system defined
as $W_{sys}=W_{1}(q_{1},p_{1})$ as function of the dimensionless
time of thermalization with the thermal environment for $\gamma=0.1$.
The inset shows the same quantities for $\gamma=0$. We considered
$(x,p_{x},y,p_{y})=(1,1,1,1)$, $(\bar{n},\bar{m})=(2,4)$ with $\bar{n}$
and $\bar{m}$ the mean number of photons of the system and thermal
environment, respectively, and $\hbar=m=\omega=1$.}

\label{fidelity_coherence}
\end{figure}

Figure (\ref{fidelity_coherence}) shows the fidelity (black solid
line) and the normalized coherence (red dashed line) for a heating
process, i.e. $\bar{n}>\bar{m}$, with $\bar{n}$ and $\bar{m}$ the
mean number of photons of the system and the thermal environment,
respectively. In order to guarantee that the environment itself generates
a Markovian evolution on the system, we show the case in which $\gamma=0$,
Fig. (\ref{fidelity_coherence})(inset), i.e. there is no coupling
of the system with any other degree of freedom. In the case with $\gamma=0.1$,
we observe the non-monotonic behavior of the fidelity and the increase
of the coherence in some time intervals.

\subsubsection*{Non-Markonian-like effect}

In Fig. (\ref{fidelity_coherence}) we note that the fidelity for
$\gamma\neq0$ presents a non-monotonic behavior during the thermalization
process with a thermal environment. This profile in the fidelity has
been recently associated to a non-Markovian dynamics on the system
characterizing an information backflow from the environment to the
system \cite{Breuer2016,Rendell2010,Rivas2010,Breuer2009}. Here the
considered thermal environment generates a Markovian dynamics on the
system, as we note from the Fig. (\ref{fidelity_coherence}) (inset)
for $\gamma=0$. Thus, any non-Markovian-like behavior arises due
to the coupling constant $\gamma\neq0$. Physically, the fidelity
is a witness of non-Markovianity and the condition is that $F(\rho_{1}(\tau),\rho_{2}(\tau))\leq F(\rho_{1}(t),\rho_{2}(t))$,
with $\rho_{1(2)}$ two arbitrary states of the system and $\tau$
and $t$ arbitrary times such that $\tau>t$ \cite{Breuer2016,Rendell2010}.
Figure (\ref{fidelity_coherence}) presents basically two time intervals
for $\gamma=0.1$ with this behavior. Furthermore, the coherence increases
during the same time intervals, evidencing a memory effect on the
system. Again, it is worth to note that this behavior is exclusively
due to the coupling $\gamma$ and does not depend on the structure
of the thermal environment. Here, a similar experiment as performed
in Ref. \cite{Cuevas2019} could be useful to simulate the non-Markovian-like
effect employing coherent (Gaussian) states in a quantum optical device.

\section{Trapped ions approach for two-coupled harmonic oscillators\label{sec:Trapped-ions-approach}}

In this section we present some well-know results that show the ability
to implement our findings in trapped ions, employing the vibrational
modes of the ion. The simplest ion trap Hamiltonian reads \cite{Ortiz2017},

\begin{equation}
H=H_{0}+H_{I},
\end{equation}
where $H_{I}$ is the interaction Hamiltonian and $H_{0}$ represents
the free dynamics of the ion and is given by

\begin{equation}
H_{0}=\frac{1}{2}\hbar\omega_{0}\sigma_{z}+\hbar\omega_{x}a_{x}a_{x}^{\dagger}+\hbar\omega_{y}a_{y}a_{y}^{\dagger},\label{Hion}
\end{equation}
with $\omega_{0}$ the transition frequency associated to the electronic
modes, $\sigma_{z}$ the Pauli matrix, $\omega_{i}$, $i=x,y$ the
natural frequencies relative to the vibrational models, which are
written in terms of the creation and annihilation operators $a_{i}$
and $a_{i}^{\dagger}$. In general, the complexity of the dynamics
depends on the well-known Lamb-Dicke parameters, $\eta_{i}=k_{0}/\sqrt{2\omega_{i}m}$,
with $k_{0}$ and $m$ the wavevector component and mass of the ion,
respectively. Let consider the Lamb-Dicke regime, $\eta_{x},\eta_{y}\ll1$.
In Ref. \cite{Ortiz2017}, the interaction Hamiltonian is expanding
in terms of the Lamb-Dicke parameters,

\begin{equation}
H_{I}=H^{(0)}+\eta_{x}H_{x}^{(1)}+\eta_{y}H_{y}^{(1)}-\eta_{x}^{2}H_{x}^{(2)}-\eta_{y}^{2}H_{y}^{(2)}-2\eta_{x}\eta_{y}^{2}H_{xy}^{(2)}+\mathcal{O}(\eta_{i}^{3}).
\end{equation}

Focusing only on the first-order terms (the other terms can be seen
in Ref. \cite{Ortiz2017}), they are given by

\begin{align}
H_{i}^{(1)} & =\frac{1}{2}\hbar\Omega\left(e^{-i(\Delta+\omega_{i})t}\sigma_{+}\hat{i}+e^{i(\Delta+\omega_{i})t}\sigma_{-}\hat{i}^{\dagger}\right)\nonumber \\
 & +\frac{1}{2}\hbar\Omega\left(e^{-i(\Delta-\omega_{i})t}\sigma_{+}\hat{i}^{\dagger}+e^{i(\Delta-\omega_{i})t}\sigma_{-}\hat{i}\right),\label{int}
\end{align}
with $\Omega$ the Rabbi frequency, $\Delta=\omega_{L}-\omega_{0}$
the radiation-atom detuning, $\omega_{L}$ the frequency of the applied
electric field, and $\sigma_{-}=|e\rangle\langle g|$ and $\sigma_{+}=|g\rangle\langle e|$
are the operators associated to the electronic transitions. This type
of interaction can be useful to experimentally simulate the first
part of our results, ie., the use of the mutual information in the
exchange of negativity between two Fock states. The interaction Hamiltonian
in Eq. (\ref{int}) allows to create the respective Fock states, once
the electronic mode can be used to create or annihilate phonons in
the vibrational modes.

On the other hand, to consider the dissipative dynamics, as in our
second result, we can use Ref. \cite{Zhong2018} and explore the generation
of cat states in trapped ions setup. Following Ref.\cite{Zhong2018}
we can write the master equation for a ion composed of one electronic
mode and two vibrational modes as,

\begin{align}
\frac{d\rho}{dt} & =-i[H_{i},\rho]+\frac{\Gamma}{2}\left(2\sigma_{-}\rho\sigma_{+}-\sigma_{+}\sigma_{-}\rho-\rho\sigma_{-}\sigma_{+}\right)\\
 & +\frac{\gamma_{x}}{2}(2a_{x}\rho a_{x}^{\dagger}-a_{x}^{\dagger}a_{x}\rho-\rho a_{x}a_{x}^{\dagger})+\frac{\gamma_{y}}{2}(2a_{y}\rho a_{y}^{\dagger}-a_{y}^{\dagger}a_{y}\rho-\rho a_{y}a_{y}^{\dagger}),
\end{align}
with $\hbar=1$, and $\Gamma$, $\gamma_{i}$ are the decay rates
to the electronic and vibrational modes, respectively, and the state
$\rho$ encodes both electronic and vibrational modes. Of course,
the coupling with the reservoir can be turned off for one vibrational
mode, thus the simulation of our results in section \ref{sec:Dissipative-dynamics}
could be experimentally implemented.

\section{Conclusions\label{sec:Conclusions}}

In this work we have considered a system composed of two-coupled harmonic
oscillators and explored some properties related to the unitary and
the dissipative dynamics. In the first case, we investigated how the
mutual information between two states allows to transfer the non-Gaussian
aspect (quantified by the negativity of the Wigner function) between
two Fock states. On the other hand, for dissipative dynamics of Gaussian
states, it was possible to note that for Gaussian states initially
with coherence, the effect of the coupling between the two subsystems
may generate some non-trivial behavior in the fidelity comparing the
state of the system and the asymptotic one, similarly to a non-Markovian-like
effect. Moreover, the same effect can be observed in the coherence
during the time evolution, with coherence revivals in specific time
intervals.

Among some experimental platforms which could be useful to experimentally
simulate our results, for instance, quantum optics and optomechanical
systems, we highlighted the trapped ions setup, where the vibrational
modes of the ion could encode the two-coupled harmonic oscillators
we investigated here, and the electronic mode could be useful to prepare
the desired states. Finally, we reinforce that the quantum information
quantities used here have been largely employed in quantum thermodynamics
protocols, showing that our results can be also applied to this field.
\begin{acknowledgments}
Jonas F. G. Santos acknowledges CAPES (Brazil), Grant No. 88882.315250/2019-01
and São Paulo Research Foundation (FAPESP), Grant No. 2019/04184-5,
for support. Carlos H. S. Vieira acknowledges CAPES (Brazil) for support.
Pedro R. Dieguez acknowledges CAPES (Brazil), Grant No. 88887.354951/2019-00.
The authors acknowledge Federal University of ABC for support.
\end{acknowledgments}

\setcounter{section}{0}
\global\long\def\thesection{Appendix \Alph{section}}

\section{Derivation of the Wigner function for two-coupled harmonic oscillators
in phase-space\label{App}}

In this appendix we derived with some details the Wigner function
for two-coupled harmonic oscillators, described in the Hamiltonian
(\ref{Hcoupled}). By introducing the creation and annihilation operators,

\begin{equation}
a_{i}=\frac{\alpha}{\sqrt{2\hbar\alpha\beta}}q_{i}+i\frac{\beta}{\sqrt{2\hbar\alpha\beta}}p_{i},
\end{equation}

\begin{equation}
a_{i}^{\dagger}=\frac{\alpha}{\sqrt{2\hbar\alpha\beta}}q_{i}-i\frac{\beta}{\sqrt{2\hbar\alpha\beta}}p_{i},
\end{equation}
with $i=1,2$, these operators $a_{i}$ and $a_{i}^{\dagger}$ satisfy
the following condition, 

\begin{equation}
\left[\hat{a}_{i},\hat{a}_{j}^{\dagger}\right]=\delta_{ij},\quad\left[\hat{a}_{i},\hat{a}_{i}^{\dagger}\right]=\left[\hat{a}_{j},\hat{a}_{j}^{\dagger}\right]=0.
\end{equation}

Now we can write the equation (\ref{Hcoupled}) as,

\begin{align}
H(q_{i},p_{i}) & =2\hbar\alpha\beta\left(a_{1}^{\dagger}a_{1}+a_{2}^{\dagger}a_{2}\right)\nonumber \\
 & -i\hbar\gamma(a_{1}a_{2}^{\dagger}-a_{2}a_{1}^{\dagger})\label{A1}
\end{align}

Note that still there exists a coupling between operators $a_{1}$
and $a_{2}$. Then, it is interesting to define the new set of creation
and annihilation operators,

\begin{equation}
A_{\pm}=\frac{1}{\sqrt{2}}\left(a_{1}\mp ia_{2}\right),\quad A_{\pm}^{\dagger}=\frac{1}{\sqrt{2}}\left(a_{1}^{\dagger}\pm ia_{2}^{\dagger}\right),
\end{equation}
with these operators obeying the following commutation relations,

\begin{align*}
\left[A_{\pm},A_{\pm}\right] & =\left[A_{\pm},A_{\mp}\right]=0\\
\left[A_{\pm}^{\dagger},A_{\pm}^{\dagger}\right] & =\left[A_{\pm}^{\dagger},\hat{A}\right]=0,\\
\left[A_{\pm},A_{\pm}^{\dagger}\right] & =\left[A_{\mp},A_{\mp}^{\dagger}\right]=1\\
\left[A_{\mp},A_{\pm}^{\dagger}\right] & =\left[A_{\pm},A_{\mp}^{\dagger}\right]=0
\end{align*}

Therefore, the Hamiltonian (\ref{A1}) is given by,

\begin{align}
H(q_{i},p_{i}) & =2\hbar\alpha\beta\left(A_{+}^{\dagger}A_{+}+A_{-}^{\dagger}A_{-}+1\right)\nonumber \\
 & -\hbar\gamma\left(A_{+}^{\dagger}A_{+}-A_{-}^{\dagger}A_{-}\right).\label{A2}
\end{align}

The next step is to rewrite the Moyal product (\ref{moyal}) in terms
of these new operators. Denoting by $A_{\pm}^{+}$ and $A_{\pm}$
as the Weyl transform of the operators $A_{\pm}^{\dagger}$ and $A_{\pm}$,
respectively, we can write the Moyal product as,

\begin{align*}
\star & =\exp\left[\frac{1}{2}\left(\overleftarrow{\partial}_{A_{+}}\overrightarrow{\partial}_{A_{+}^{+}}-\overleftarrow{\partial}_{A_{+}^{+}}\overrightarrow{\partial}_{A_{+}}\right)\right]\\
 & \times\text{exp}\left[\frac{1}{2}\left(\overleftarrow{\partial}_{A_{-}}\overrightarrow{\partial}_{A_{-}^{+}}-\overleftarrow{\partial}_{A_{-}^{+}}\overrightarrow{\partial}_{A_{-}}\right)\right],
\end{align*}
with $A_{+}^{+}\star A_{+}=A_{+}^{+}A_{+}-1/2.$

Defining new variables $\eta_{1}=A_{+}^{+}A_{+}$ and $\eta_{2}=A_{-}^{+}A_{-}$,
the Weyl transform of the Hamiltonian (\ref{A2}) reads,

\begin{equation}
H^{W}(\eta_{1},\eta_{2})=(2\hbar\alpha\beta-\hbar\gamma)\eta_{1}+(2\hbar\alpha\beta+\hbar\gamma)\eta_{2}.
\end{equation}

In addition, applying the so-called stargenvalue equation (\ref{eigenvalueeq})
for the system,

\begin{equation}
H^{W}(\eta_{1},\eta_{2})\star W(\eta_{1},\eta_{2})=EW(\eta_{1},\eta_{2}),
\end{equation}
we have that,

\begin{align*}
H^{W}(\eta_{1},\eta_{2})\star W(\eta_{1},\eta_{2}) & =[(2\hbar\alpha\beta-\hbar\gamma)\eta_{1}+(2\hbar\alpha\beta+\hbar\gamma)\eta_{2}\\
 & +\frac{1}{2}(2\hbar\alpha\beta-\hbar\gamma)+\left(A_{+}^{+}\frac{\partial}{\partial A_{+}^{+}}-A_{+}\frac{\partial}{\partial A_{+}}\right)\\
 & +\frac{1}{2}(2\hbar\alpha\beta+\hbar\gamma)+\left(A_{-}^{+}\frac{\partial}{\partial A_{-}^{+}}-A_{-}\frac{\partial}{\partial A_{-}}\right)\\
 & -\frac{1}{4}(2\hbar\alpha\beta-\hbar\gamma)-\frac{\partial^{2}}{\partial A_{+}\partial A_{+}^{+}}\\
 & -\frac{1}{4}(2\hbar\alpha\beta+\hbar\gamma)-\frac{\partial^{2}}{\partial A_{-}\partial A_{-}^{+}}]\\
 & =EW(\eta_{1},\eta_{2}).
\end{align*}

After some mathematical manipulations we obtain two uncoupled differential
equations given by,

\begin{align}
\left[\eta_{1(2)}-\frac{1}{4}\left(\frac{\partial}{\partial\eta_{1(2)}}+\eta_{1(2)}\frac{\partial^{2}}{\partial\eta_{1(2)}^{2}}\right)-E_{1(2)}\right]\nonumber \\
\times\chi_{1(2)}(\eta_{1(2)}) & =0\label{A3}
\end{align}

The explicit solutions of Eqs. (\ref{A3}) can be written in terms
of the Laguerre polynomials,

\begin{equation}
\chi_{1(2)}(\eta_{1(2)})=(-1)^{n_{1(2)}}\exp\left[-2\chi_{1(2)}\right]L_{n_{1(2)}}\left[4\chi_{1(2)}\right],
\end{equation}
with $n_{1(2)}$ non-negative integers and $n_{1(2)}=(E_{1(2)}-1/2)$,
resulting in,

\begin{equation}
E_{n_{1},n_{2}}=2\hbar\alpha\beta\left(n_{1}+n_{2}+1\right)+\hbar\gamma\left(n_{1}-n_{2}\right).
\end{equation}
Finally, writing as function of canonical phase-space variables, the
Wigner function in (\ref{WignerA}) is obtained.


\begin{thebibliography}{10}
\bibitem{Pleni02}A. Bermudez, T. Schaetz, and M. B. Plenio, Dissipation-Assisted
Quantum Information Processing with Trapped Ions, Phys. Rev. Lett.
$\mathbf{110}$, 110502 (2013).

\bibitem{Weekbrook2014}K. Marshall and C. Weedbrook, Device-independent
quantum cryptography for continuous variables, Phys. Rev. A \textbf{90},
042311 (2014).

\bibitem{Pirandola2017}P. Papanastasiou, C. Ottaviani, and S. Pirandola,
Finite-size analysis of measurement-device-independent quantum cryptography
with continuous variables, Phys. Rev. A\textbf{ 96}, 042332 (2017).

\bibitem{Pirandola2015}C. Ottaviani, S. Mancini, and S. Pirandola,
Two-way Gaussian quantum cryptography against coherent attacks in
direct reconciliation, Phys. Rev. A \textbf{92}, 062323 (2015).

\bibitem{Cirac2019}M. Perarnau-Llobet, A. González-Tudela, J. I.
Cirac, Multimode Fock states with large photon number: effective descriptions
and applications in quantum metrology, arXiv:1910.03323.

\bibitem{Adesso2017}Pietro Liuzzo-Scorpo, Andrea Mari, Vittorio Giovannetti,
and Gerardo Adesso, Optimal Continuous Variable Quantum Teleportation
with Limited Resources, Phys. Rev. Lett. \textbf{119}, 120503 (2017).

\bibitem{Camati2019}P. A. Camati, J. F. G. Santos, and R. M. Serra,
Coherence effects in the performance of the quantum Otto heat engine,
Phys. Rev. A \textbf{99}, 062103 (2019).

\bibitem{Kosloff2019}R. Dann and R. Kosloff, Quantum Signatures in
the Quantum Carnot Cycle, arXiv:1906.0694.

\bibitem{Shanhe2019}S.Su, W. Shen, J. Du, and J. Chen, Coherence
induced work in quantum heat engines with Larmor precession, arXiv:1908.06443.

\bibitem{Zureck01}W. H. Zurek, Reversibility and Stability of Information
Processing Systems, Phys. Rev. Lett. $\mathbf{53}$, 391 (1984).

\bibitem{Supre2019}F. Arute, K. Arya, R. Babbush, et al. Quantum
supremacy using a programmable superconducting processor. Nature \textbf{574},
505-510 (2019).

\bibitem{Silva2015}Z. Gedik, I. A. Silva, B. Çakmak, G. Karpat, E.
L. G. Vidoto, D. O. Soares-Pinto, E. R. deAzevedo and F. F. Fanchini,
Computational speed-up with a single qudit, Sci. Rep. \textbf{5},
14671 (2015).

\bibitem{Partovi2008}M. H. Partovi, Entanglement versus stosszahlansatz:
disappearance of the thermodynamic arrow in a high-correlation environment.
Phys. Rev. E \textbf{77}, 021110 (2008).

\bibitem{Jevtic2012}S. Jevtic, D. Jennings, and T. Rudolph, Maximally
and minimally correlated states attainable within a closed evolving
system. Phys. Rev. Lett. \textbf{108}, 110403 (2012).

\bibitem{Serra2019}K. Micadei, J.P.S. Peterson, A.M. Souza et al.
Reversing the direction of heat flow using quantum correlations. Nat
Commun \textbf{10}, 2456 (2019).

\bibitem{Maniscalco2018}S. Hamedani Raja, M. Borrelli, R. Schmidt,
J. P. Pekola, S. Maniscalco, Thermodynamic fingerprints of non-Markovianity,
Phys. Rev. A \textbf{97}, 032133 (2018).

\bibitem{Maniscalco2017}M. Cianciaruso, S. Maniscalco, and G. Adesso,
Role of non-Markovianity and backflow of information in the speed
of quantum evolution, Phys. Rev. A \textbf{96}, 012105 (2017).

\bibitem{Camati2020}P. A. Camati, J. F. G. Santos, and R. M. Serra,
Employing Non-Markovian effects to improve the performance of a quantum
Otto refrigerator. In preparation.

\bibitem{Feig2019}W. Ge, B. C. Sawyer, J. W. Britton, K. Jacobs,
J. J. Bollinger, and M. Foss-Feig, Trapped Ion Quantum Information
Processing with Squeezed Phonons, Phys. Rev. Lett. \textbf{122}, 030501
(2019).

\bibitem{Giovannetti01}A. Carlini, A. Mari, and V. Giovannetti, Time-optimal
thermalization of single-mode Gaussian states, Phys. Rev. A $\mathbf{90}$,
052324 (2014).

\bibitem{Serafini2018}F. Albarelli, U. S.-Bennett, and A. Serafini,
Locally optimal control of continuous-variable entanglement, Phys.
Rev. A\textbf{ 98}, 062312 (2018).

\bibitem{Su2019}D.Su, C. R. Myers, and K. K. Sabapathy, Conversion
of Gaussian states to non-Gaussian states using photon-number-resolving
detectors, Phys. Rev. A \textbf{100}, 052301 (2019).

\bibitem{Adesso}G. Adesso, S. Ragy, and A. R. Lee, Continuous variable
quantum information: Gaussian states and beyond, Open Syst. Inf. Dyn.$\mathbf{21}$,
1440001 (2014).

\bibitem{Weedbrook2012}C. Weedbrook, S. Pirandola, R. García-Patrón,
N. J. Cerf, T. C. Ralph, J. H. Shapiro, and S. Lloyd, Gaussian quantum
information, Rev. Mod. Phys. \textbf{84}, 621 (2012).

\bibitem{Petruccione01}H. P. Breuer and F. Petruccione, The Theory
of Open Quantum Systems, (Oxdord University Press, Oxford, 2002).

\bibitem{Nielsen}M. A. Nielsen and I. L. Chuang, Quantum Computation
and Quantum Information, (Cambridge University Press, Cambridge, 2010).

\bibitem{Quantumopticsbook}M. O. Scully and M. S. Zubairy, Quantum
optics, (Cambridge University Press, Cambridge, 2012).

\bibitem{Wineland2003}D. Leibfried, R. Blatt, C. Monroe, and D. Wineland,
Quantum dynamics of single trapped ions, Rev. Mod. Phys. \textbf{75},
281 (2003).

\bibitem{Sage2019}C. D. Bruzewicz, J. Chiaverini, R.t McConnell,
and J.M. Sage, Trapped-Ion Quantum Computing: Progress and Challenges,
arXiv:1904.04178.

\bibitem{Biedermann2015}G. W. Biedermann, X. Wu, L. Deslauriers,
S. Roy, C. Mahadeswaraswamy, and M. A. Kasevich, Testing gravity with
cold-atom interferometers, Phys. Rev. A \textbf{91}, 033629 (2015).

\bibitem{Han2018}Y. Han, C. Zhu, X. Huang, Y. Yang, Electromagnetic
control of nonclassicality in cavity QED system, Physical Review A
\textbf{98}, 033828 (2018).

\bibitem{Werlang2018}A. V. Dodonov, D. Valente, T. Werlang, Quantum
power boost in a nonstationary cavity-QED quantum heat engine, J.
Phys. A: Math. Theor. \textbf{51}, 365302 (2018).

\bibitem{Zachosbook}C. K Zachos, D. B Fairlie, and T. L Curtright,
Quantum Mechanics in Phase Space: An Overview with Selected Papers,
(World Scientific Pub. Co. Inc., Singapure, 2005).

\bibitem{Wigner1932}E. Wigner, On the Quantum Correction For Thermodynamic
Equilibrium, Phys. Rev. \textbf{40}, 749 (1932).

\bibitem{Case2008}W. B. Case, Wigner functions and Weyl transforms
for pedestrians, Am. J. Phys. 76, 937 (2008).

\bibitem{Chou2018} K.S. Chou, J.Z. Blumoff, C.S. Wang et al. Deterministic
teleportation of a quantum gate between two logical qubits. Nature
\textbf{561}, 368 (2018).

\bibitem{Santos2015}J. F. G. Santos, A. E. Bernardini, and C. Bastos,
Probing phase-space noncommutativity through quantum mechanics and
thermodynamics of free particles and quantum rotors, Physica A $\mathbf{438}$,
340 (2015)

\bibitem{Santos2016}J. F. G. Santos and A. E. Bernardini, Gaussian
fidelity distorted by external fields, Physica A $\mathbf{445}$,
75 (2016).

\bibitem{Bernardini01}A. E. Bernardini and O. Bertolami, Probing
phase-space noncommutativity through quantum beating, missing information
and the thermodynamic limit, Phys. Rev. A $\mathbf{88}$, 012101 (2013).

\bibitem{Rosenbaum2006}M. Rosenbaum and J. D. Vergara, The \ensuremath{\star}-value
equation and Wigner distributions in noncommutative Heisenberg algebras,
Gen. Rel. Grav. \textbf{38}, 607 (2006).

\bibitem{Twamley1997}J. Steinbach, J. Twamley, and P. L. Knight,
Engineering two-mode interactions in ion traps, Phys. Rev. A \textbf{56},
4815 (1997).

\bibitem{Wineland1996}C. Monroe, D. M. Meekhof, B. E. King, and D.
J. Wineland, A \textquotedblleft Schrödinger Cat\textquotedblright{}
Superposition State of an Atom, Science \textbf{272}, 5265 (1996).

\bibitem{Zhong2018}Z.-R. Zhong, X.-J. Huang, Z.-B.Yang, L.-T. Shen,
and S.-B. Zheng, Generation and stabilization of entangled coherent
states for the vibrational modes of a trapped ion, Phys. Rev. A \textbf{98},
032311 (2018).

\bibitem{Ortiz2017}L. Ortiz-Gutiérrez, B. Gabrielly, L. F. Muñoz,
K. T. Pereira, J. G. Filgueiras, and A. S. Villar, Continuous variables
quantum computation over the vibrational modes of a single trapped
ion, Optics Communications \textbf{397}, (2017).

\bibitem{Paris2018}F. Albarelli, M. G. Genoni, and M. G. A. Paris,
Resource theory of quantum non-Gaussianity and Wigner negativity,
Phys. Rev. A \textbf{98}, 052350 (2018).

\bibitem{Deering2015}K.-P. Marzlin and S. Deering, The Moyal equation
for open quantum systems, J. Phys. A: Math. Theor. \textbf{48}, 205301
(2015).

\bibitem{Mandredi2000}G. Manfredi, and M. R. Feix, Entropy and Wigner
functions, Phys. Rev. E \textbf{62}, 4665 (2000).

\bibitem{Kenfack2014}A. Kenfack and K. Zyczkowski, Negativity of
the Wigner function as an indicator of non-classicality, J. Opt. B:
Quantum Semiclass. Opt. \textbf{6} 396 (2004).

\bibitem{Rundle2017}R. P. Rundle, P. W. Mills, Todd Tilma, J. H.
Samson, and M. J. Everitt, Simple procedure for phase-space measurement
and entanglement validation, Phys. Rev. A Phys. Rev. A 96, 022117
(2017).

\bibitem{Treps2017}M.Walschaers, C. Fabre, V. Parigi, and N. Treps,
Entanglement and Wigner Function Negativity of Multimode Non-Gaussian
States, Phys. Rev. Lett. 119, 183601 (2017).

\bibitem{Svozil=0000EDk2018}I. I. Arkhipov, A.Barasi\'{n}ski, and
J. Svozilík, Negativity volume of the generalized Wigner function
as an entanglement witness for hybrid bipartite states, Sci. Rep.
\textbf{8}, 16955 (2018).

\bibitem{Haroche2000}G. Nogues, A. Rauschenbeutel, S. Osnaghi, P.
Bertet, M. Brune, J. M. Raimond, S. Haroche, L. G. Lutterbach, and
L. Davidovich, Measurement of a negative value for the Wigner function
of radiation, Phys. Rev. A \textbf{62}, 054101 (2000).

\bibitem{Okay2014}R.Raussendorf, D. E. Browne, N. Delfosse, C. Okay,
and J. Bermejo-Vega, Contextuality and Wigner-function negativity
in qubit quantum computation, Phys. Rev. A \textbf{95}, 052334 (2014).

\bibitem{Hiroshima2007}X.-B. Wang, T. Hiroshima, A. Tomita, and M.
Hayashi, Quantum information with Gaussian states, Physics Reports
\textbf{448}, (2007).

\bibitem{Serafini}A. Serafini, Quantum Continuous Variables. A primer
of Theoretical Methods, (CRC Press, Boca Raton, 2017).

\bibitem{Holevo}A. S. Holevo, Some statistical problems for quantum
Gaussian states, IEEE Trans. Inf. Theory $\mathbf{21}$, 533 (1975).

\bibitem{Scutaru}H. Scutaru, Fidelity for displaced squeezed states
and the oscillator semigroup, J. Phys. A $\mathbf{31}$, 3659 (1998).

\bibitem{Hage2007}J. DiGuglielmo, B. Hage, A. Franzen, J. Fiurášek,
and Roman Schnabel, Experimental characterization of Gaussian quantum-communication
channels, Phys. Rev. A \textbf{76}, 012323 (2007).

\bibitem{Winter2016}A. Winter and D. Yang, Operational Resource Theory
of Coherence, Phys. Rev. Lett. \textbf{116}, 120404 (2016).

\bibitem{Plenio2014}T. Baumgratz, M. Cramer, and M. B. Plenio, Quantifying
coherence, Phys. Rev. Lett. 113, 140401 (2014).

\bibitem{Feldmann2006}T. Feldmann and R. Kosloff, Quantum lubrication:
Suppression of friction in a first-principles four-stroke heat engine,
Phys. Rev. E \textbf{73}, 025107(R) (2006).

\bibitem{Rezek2006}Y. Rezek and R. Kosloff, Irreversible performance
of a quantum harmonic heat engine, New J. Phys. \textbf{8}, 83 (2006).

\bibitem{Xu2016}J. Xu, Quantifying coherence of Gaussian states,
Phys. Rev. A $\boldsymbol{93}$, 032111 (2016).

\bibitem{Giovanneti01}A. Carlini, A. Mari, and V. Giovanneti, Time-Optimal
Thermalization of Single-Mode Gaussian States, Phys. Rev. A $\mathbf{90}$,
052324 (2014).

\bibitem{Sahoo2007}A. M. Jayannavar and Mamata Sahoo, Charged particle
in a magnetic field: Jarzynski equality, Phys. Rev. A \textbf{75},
032102 (2007).

\bibitem{Munos2014}E. Muñoz and F. J. Peña, Magnetically driven quantum
heat engine Phys. Rev. E \textbf{89}, 052107 (2014).

\bibitem{Santos2017}J. F. G. Santos, A. E. Bernardini, Quantum engines
and the range of the second law of thermodynamics in the noncommutative
phase-space, Eur. Phys. J. Plus \textbf{132}, 260 (2017).

\bibitem{Rivas2010}Á. Rivas, S. F. Huelga, and M. B. Plenio, Entanglement
and Non-Markovianity of Quantum Evolutions, Phys. Rev. Lett. $\mathbf{105}$,
050403 (2010).

\bibitem{Breuer2009}H.-P. Breuer, E.-M. Laine, and J. Piilo, Measure
for the Degree of Non-Markovian Behavior of Quantum Processes in Open
Systems, Phys. Rev. Lett. $\mathbf{103}$, 210401 (2009).

\bibitem{Breuer2016}H.-P. Breuer, E.-M. Laine, J. Piilo, and B. Vacchini,
Colloquium: Non-Markovian dynamics in open quantum systems, Rev. Mod.
Phys. $\mathbf{88}$, 021002 (2016).

\bibitem{Rendell2010}A. K. Rajagopal, A. R. Usha Devi, and R. W.
Rendell, Kraus representation of quantum evolution and fidelity as
manifestations of Markovian and non-Markovian forms A., Phys. Rev.
A $\mathbf{82}$, 042107 (2010).

\bibitem{Cuevas2019}Á. Cuevas, A. Geraldi, C. Liorni et al. All-optical
implementation of collision-based evolutions of open quantum systems.
Sci Rep \textbf{9}, 3205 (2019).
\end{thebibliography}
\end{document}